\documentclass[journal]{IEEEtran}
\ifCLASSINFOpdf
\else
\fi
\usepackage{graphicx}
\usepackage{amsmath}
\usepackage{amssymb}    
\usepackage{booktabs}
\usepackage[numbers]{natbib}
\let\labelindent\relax
\usepackage{enumitem} 
\usepackage{multirow}
\usepackage{calc} 
\usepackage[dvipsnames]{xcolor}   
\usepackage{soul}                 

\begin{document}
%
\title{OrthoInsight: Rib Fracture Diagnosis and Report Generation Based on Multi-Modal Large Models}
%
%
%

\author{Ningyong~Wu,
        Jiangbo~Zhang,
        Wenhong~Zhao, 
        Jinzhi~Wang,
        Chenzhan~Yu,
        Zhigang~Xiu,
        Duwei~Dai,
        Ziyu~Xu,
        and Yongli~Yang
\thanks{N. Wu and W. Zhao are with the Organizational Management Department, 
School of Management, Xi’an Jiaotong University, Xian, Shanxi, China. 
(Corresponding author: Wenhong Zhao, e-mail: jdzhao@xjtu.edu.cn; 
homepage: http://gr.xjtu.edu.cn/web/jdzhao).}
\thanks{N. Wu, C. Yu, and Z. Xiu are with West China Longquan Hospital, 
Sichuan University, Chengdu, Sichuan, China.}
\thanks{J. Zhang is with the School of Electronic Science and Engineering, 
Xi’an Jiaotong University, No.~28, Xianning West Road, Beilin District, 
Xi’an 710049, China (e-mail: 4522253009@stu.xjtu.edu.cn).}
\thanks{J. Wang is with the Systems Engineering Institute, 
Xi’an Jiaotong University, Xian, Shanxi, China.(e-mail: wangjz5515@stu.xjtu.edu.cn).}

\thanks{D. Dai is with the Institute of Medical Artificial Intelligence, 
the Second Affiliated Hospital of Xi’an Jiaotong University, 
Xian, Shanxi, China.}
\thanks{Z. Xu is a Ph.D. candidate with the School of Human Settlements and Civil Engineering,
Xi’an Jiaotong University, Xi’an, China. 
(e-mail: Xzy365@stu.xjtu.edu.cn).}%
\thanks{Y. Yang is a Ph.D. candidate with the School of Life Science and Technology,
Xi’an Jiaotong University, Xi’an, China. 
(e-mail: yyl123456@stu.xjtu.edu.cn).}%
}
\maketitle

\begin{abstract}

The growing volume of medical imaging data has increased the need for automated diagnostic tools, especially for musculoskeletal injuries like rib fractures commonly assessed on CT. Manual interpretation is time-consuming and error-prone; moreover, subtle fractures, multi-site injuries, and the demand for clinically actionable recommendations make both detection and reporting challenging in routine practice. To address these issues, we propose OrthoInsight, a multimodal deep learning framework for rib fracture diagnosis and report generation. It integrates a YOLOv9 model for fracture detection, a medical knowledge graph for retrieving clinical context, and a fine-tuned LLaVA language model for generating diagnostic reports. OrthoInsight combines visual features from CT images with expert textual data to deliver clinically useful outputs. Evaluated on 28,675 annotated CT images and expert reports, it achieves high performance across Diagnostic Accuracy, Content Completeness, Logical Coherence, and Clinical Guidance Value, with an average score of 4.28, outperforming models like GPT-4 and Claude-3. This study demonstrates the potential of multimodal learning in transforming medical image analysis and providing effective support for radiologists.

\end{abstract}


%
\IEEEpeerreviewmaketitle

\section{Introduction}
%
%
%
%
\IEEEPARstart{T}{he} advancement of medical imaging technologies has greatly improved the ability to detect and diagnose a wide range of medical conditions, particularly musculoskeletal injuries such as rib fractures \cite{Chapman_2016}. Rib fractures are a common injury that can result from trauma or accidents and are often diagnosed through the use of computed tomography (CT) images \cite{LIU2022283}. While traditional methods involving manual interpretation of CT images by radiologists have been widely employed, this process is inherently time-consuming, requires a high level of expertise, and is subject to human error \cite{FLORY2024152}. As the volume of medical imaging data continues to grow, it has become increasingly difficult for radiologists to efficiently and accurately interpret every image. Recent studies demonstrate that AI-assisted workflows can significantly improve diagnostic sensitivity while maintaining high specificity \cite{Sun_2025}. This has created a significant need for automated systems that can assist clinicians in making accurate diagnoses more quickly and effectively.

\begin{figure}
    \centering
    \includegraphics[width=1\linewidth]{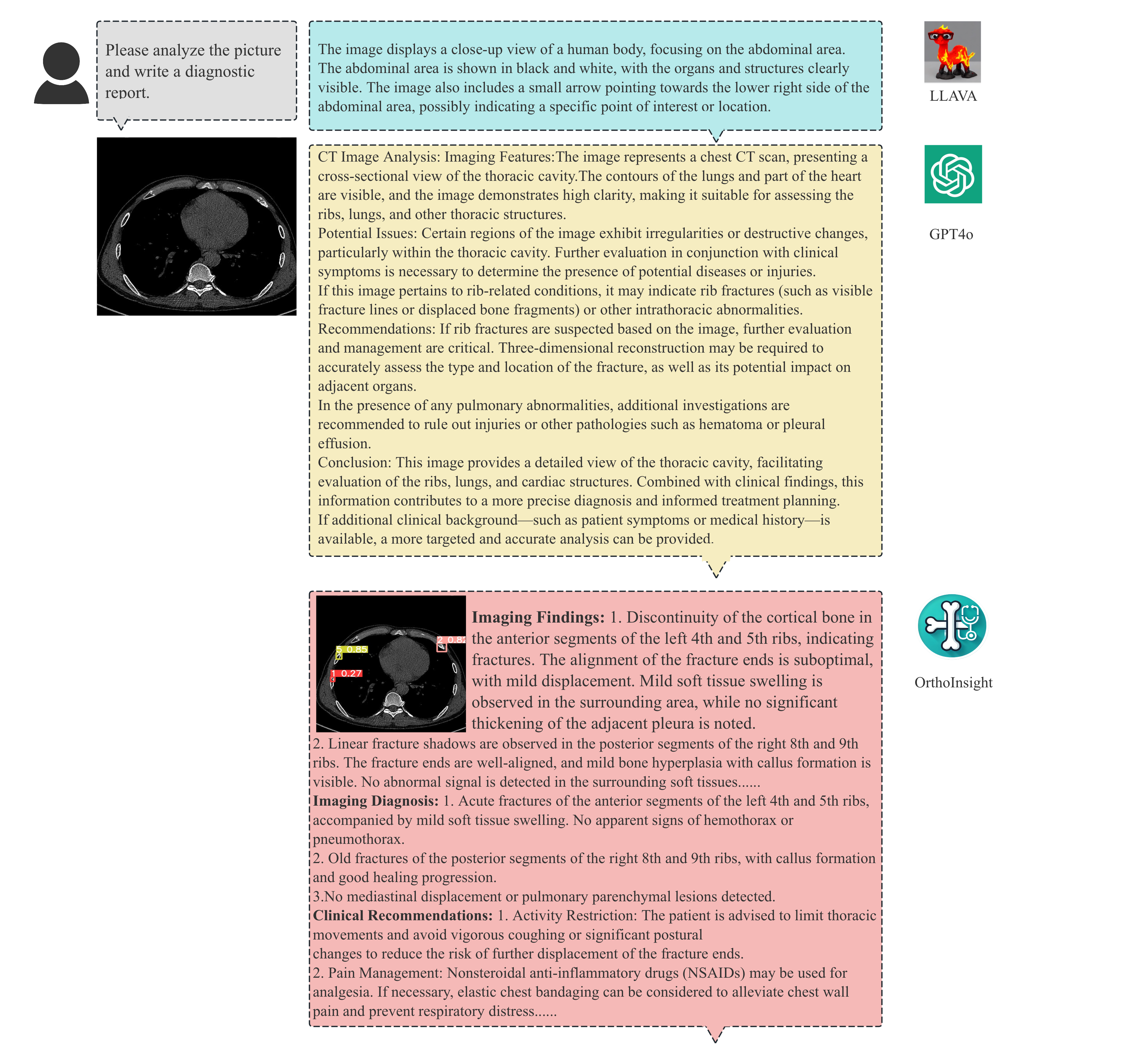}
    \caption{Comparison of LLaVA, GPT-4V and OrthoInsight.}
    \label{FIG:1}
\end{figure}

In recent years, the development of artificial intelligence (AI) and deep learning techniques has shown considerable promise in revolutionizing medical image analysis. Among these, convolutional neural networks (CNNs \cite{LeCun_1998}) and more advanced models, such as YOLO (You Only Look Once \cite{Redmon_2016}), have become widely used for image detection tasks due to their ability to quickly identify objects within images with high accuracy. For example, improved YOLO variants achieve a mean average precision (mAP50) of 0.972 in rib fracture detection\cite{Li_2025}. However, while these models excel at detecting visual features, the integration of contextual and domain-specific knowledge remains a critical challenge. In clinical practice, a diagnosis is often more than just a diagnosis of an issue: It requires understanding the underlying causes, possible treatment options, and the broader context of the patient’s health. This gap in understanding has led to the emergence of multimodal learning approaches, which combine both image data and auxiliary text (such as clinical reports or medical knowledge) to generate more comprehensive output \cite{li2023llavamedtraininglargelanguageandvision}.

The application of multimodal learning to medical imaging holds significant potential in improving diagnostic accuracy and efficiency. By combining visual information from CT images with textual data from expert reports and established medical knowledge, multi-modal models can generate detailed diagnostic insights that are not only more accurate but also more clinically actionable. For example, multimodal models can detect rib fractures, classify their type and severity, and generate automated diagnostic reports that include visual and textual descriptions, helping clinicians make informed decisions \cite{guo2024llavaultralargechineselanguage}. 

In this study, we propose a novel multimodal deep learning framework, OrthoInsight, specifically designed for rib-fracture diagnosis from CT images. The framework integrates a YOLOv9-based model \cite{wang2024yolov9learningwantlearn} to localize rib fractures and predict fracture-related attributes, a medical knowledge graph to retrieve relevant clinical context, and a pre-trained multimodal language model (LLaVA  \cite{liu2023visualinstructiontuning}) to generate detailed and clinically useful diagnostic reports. By coupling visual evidence with domain knowledge, OrthoInsight aims to reduce missed detections in subtle or multi-site fractures and alleviate the reporting burden by producing actionable recommendations.

The primary contributions of this work are as follows:
\begin{itemize}
    \item End-to-end knowledge-grounded pipeline: We propose OrthoInsight, an integrated framework that bridges rib-fracture detection and report generation by combining YOLOv9-based localization/classification with multimodal LLM report writing, reducing the gap between image findings and clinically actionable narratives.
    \item Clinical knowledge integration: We curate and leverage an orthopedic knowledge graph to retrieve etiology, management options, and complication-monitoring suggestions, enabling the generated reports to be evidence-aware and more clinically informative than vision-only outputs.
    \item Comprehensive evaluation: We build and evaluate on a large-scale dataset (28,675 CT images with expert reports) and adopt clinically oriented metrics (DA/CC/LCC/CGV) to assess both diagnostic correctness and the practical value of generated recommendations.
\end{itemize}

The remainder of this paper is organized as follows. Section II reviews related work. Section III formulates the task and metrics. Section IV introduces dataset and knowledge-graph construction. Section V describes the proposed methodology, and Section VI presents experimental setups. Section VII reports the main results and ablations, followed by a qualitative case study in Section VIII. Finally, Section IX concludes the paper and Section X provides the ethics statement.
\section{Related Work}

\subsection{Applications of Large Models in the Medical Field}

The application of large-scale pre-trained models (e.g., GPT-4, Med-PaLM) in the medical field has made significant progress, particularly in medical text analysis and clinical decision support. In recent years, large language models (LLMs) based on natural language processing (NLP) have been widely used to handle tasks such as electronic health records (EHR), medical record summarization, and medical question-answering. LLMs, with their powerful text comprehension abilities, can automate the processing of large volumes of medical text and provide accurate clinical recommendations. For example, GPT-4 and Med-PaLM have made breakthroughs in medical question-answering tasks, especially in tests modeled after the United States Medical Licensing Examination (USMLE), achieving high scores (e.g., 86. 4\% precision for GPT-4 and 67.6\% for Med-PaLM) and demonstrating their strong capabilities in understanding complex medical issues and clinical scenarios \cite{Singhal_2023} \cite{nori2023capabilitiesgpt4medicalchallenge}. Recent studies further validate that domain-specific LLMs such as Meditron-70B achieve performance comparable to general-purpose models while requiring fewer computational resources \cite{lou2024poweroptimizationintegratedactive} \cite{VanVeen_2023}.

In terms of question-answering capabilities, these models can also be applied to tasks such as medical text summarization and medical record information extraction. By training on large amounts of clinical text, LLMs can generate high-quality medical summaries, which hold great potential for reducing clinical documentation workload and improving efficiency. Research has shown that LLMs fine-tuned in the medical domain can generate summaries that are as accurate as, or even more accurate than, those written by human doctors (87\% of clinicians rated LLM-generated summaries as equivalent or superior to human-written ones) \cite{Adams_2023} \cite{VanVeen_2023}. For instance, a hybrid model combining BART and rule-based post-editing improved radiology report summarization accuracy by 12\% compared to standalone LLMs \cite{Shakil_2024}. The applications of these models are not limited to extracting critical information to support clinical decision-making but also extend to providing personalized treatment plans for patients. As medical data continues to grow, systems that combine LLMs with medical knowledge graphs are gradually being deployed, assisting in diagnosis and decision-making by integrating multi-layered knowledge \cite{CHEN2025126215} \cite{zuo2025kg4diagnosishierarchicalmultiagentllm}. A notable example is the use of dynamic knowledge graphs updated in real-time with EHR data, which enhanced treatment recommendation relevance by 23\% in oncology cases \cite{Hasan_2020}. Ethical and regulatory challenges remain critical. Studies highlight that over 30\% of LLM-generated recommendations may contain biases related to patient demographics, emphasizing the need for rigorous fairness audits \cite{Yang_2024}.

\subsection{Applications of Multimodal Large Models in Medical Image Analysis}

The use of multimodal large models (e.g., LLaVA, Med-Alpaca) in medical image analysis has become an important area of current research \cite{li2023llavamedtraininglargelanguageandvision} \cite{han2025medalpacaopensourcecollection}. Traditional medical image analysis methods typically rely solely on image data, often overlooking critical information from text data, such as radiology reports and clinical notes \cite{chen2024advancinghighresolutionvisionlanguage}. However, medical images often provide only local visual information, whereas a patient’s medical history, clinical symptoms, and the physician’s initial diagnosis are often key to making an accurate diagnosis \cite{li2023llavamedtraininglargelanguageandvision}. To address this, multimodal large models combine image data and text data, significantly improving the accuracy of automated diagnosis \cite{he2025pefomedparameterefficientfinetuning}.

In the field of medical imaging, multimodal large models primarily process both image and related text data through joint training of visual and language models. For example, the LLaVA-Med model enhances the understanding of medical images by leveraging multimodal learning. Trained on biomedical image and description datasets, the model can simultaneously process medical images and clinical text, generating more accurate image analysis reports. Moreover, approaches based on visual-language models (VLMs), such as GPT-4V, can recognize visual features in medical images and use language processing techniques to interpret and analyze these features\cite{yang2023dawnlmmspreliminaryexplorations} \cite{chen2024advancinghighresolutionvisionlanguage}.

Furthermore, combining knowledge graphs with large models further enhances the quality of image analysis \cite{nguyen2024logramedlongcontextmultigraph}. Integrating systems with medical ontologies like LS and SNOMED CT enables a better understanding of the clinical context behind medical images and provides doctors with more precise and interpretable diagnostic suggestions \cite{Chang_2024}. This integration makes large models not only perform well in image classification and segmentation tasks but also excel in disease diagnosis, treatment recommendations, and follow-up care \cite{Yang_2025}.

In summary, prior studies have shown promising progress in (i) rib-fracture detection on CT and (ii) medical report generation with large language models. However, many pipelines still treat detection and report writing as separate stages and lack explicit orthopedic knowledge grounding, which is critical for producing clinically actionable recommendations (e.g., follow-up planning and complication monitoring). In addition, evaluation is often dominated by generic text-similarity metrics that may not fully reflect clinical usefulness. These limitations motivate OrthoInsight, which unifies fracture localization, knowledge retrieval, and multimodal report generation under clinically oriented evaluation criteria.

\section{Task Setup}
This task generates a detailed diagnostic analysis report from CT rib images. The input consists of a CT rib image and a prompt to guide the model in generating the report. The output is a diagnostic report containing image findings, image diagnosis, and clinical recommendations. The mathematical formulation is as follows:

\begin{equation}
R = \text{Gen}(\text{Fuse}(\text{ImgFeat}(I), \text{TxtEnc}(P))\ = (T, C, S)
\end{equation}

In this equation, \( R \) represents the generated report, \( \text{ImgFeat}(I) \) refers to the image feature extractor applied to the CT rib image \( I \), and \( \text{TxtEnc}(P) \) refers to the text encoder used to process the prompt \( P \). The function \( \text{Fuse} \) combines the image features and text encoding. \( \text{GenReport} \) is the report generation function. The generated report \( R \) consists of three main components: \( T \) represents image findings, \( C \) represents image diagnosis, and \( S \) provides clinical recommendations.
\subsection{Evaluation metrics}

In the general domain, evaluation metrics typically focus on text similarity, which may not fully address the specific requirements of the medical field, such as the precise use of medical terminology, the clinical practicality of treatment recommendations, and the logical rigor of the report. Therefore, to ensure that diagnostic reports effectively support clinical practice, specialized evaluation metrics must be established. Based on the professional opinions of orthopedic experts, practical application needs, and relevant evaluation standards from healthcare \cite{tiffin2011evaluating} \cite{zhang2023huatuogpttaminglanguagemodel}, education \cite{ghafourian2023readability}, and software engineering \cite{karunaratne2022review} fields, we have identified four core characteristics that diagnostic reports should possess to ensure their reliability and clinical applicability in medical settings:
\begin{table*}[h!]
\centering
\caption{Detailed scoring criteria for metrics}
\label{tabel1}
\begin{tabular}{p{2cm}p{12cm}}
\toprule
\textbf{Metrics} & \textbf{Scoring details} \\ \midrule
\textbf{Diagnostic Accuracy (DA)} & 
\begin{tabular}[c]{p{12cm}}5 points: Accurately identifies the type, location, and severity of the rib fracture. The diagnosis is correct and provides sufficient support for clinical decision-making.\\ 4 points: The diagnosis is accurate, recognizing the type, location, and severity of the fracture, but the description may be slightly brief or lack some details.\\ 3 points: Able to identify the type or location of the fracture but lacks a description of the severity, or the information is unclear.\\ 2 points: The diagnosis is vague or incomplete, failing to identify the fracture type, location, or severity accurately, and may lack crucial information.\\ 1 point: The diagnosis is incorrect or missing, failing to identify the fracture type, location, or severity, which impacts clinical treatment decisions.\end{tabular} \\ \midrule
\textbf{Content Completeness (CC)} &
\begin{tabular}{p{12cm}}5 points: The report includes all necessary orthopedic information, including imaging findings, diagnosis, treatment recommendations, and follow-up plans. The content is comprehensive and detailed.\\ 4 points: The report includes most of the key information, with a few minor details missing, but still provides effective support for clinical decision-making.\\ 3 points: The report lacks some essential information, affecting the diagnosis and treatment decisions, though some conclusions can still be inferred from other parts.\\ 2 points: The report is incomplete, missing key details regarding the diagnosis or treatment plan, which affects clinical judgment.\\ 1 point: The report severely lacks critical information and cannot effectively support clinical decisions.\end{tabular} \\ \midrule
\textbf{Logical Coherence and Consistency (LCC)} &
\begin{tabular}[c]{p{12cm}}5 points: The report has a clear structure with well-organized information, following a logical sequence such as imaging findings $\rightarrow$ diagnosis $\rightarrow$ clinical recommendations, making it easy to understand.\\ 4 points: The report is generally well-structured with relatively coherent information. Minor structural issues may exist, but it remains understandable overall.\\ 3 points: The report’s structure is somewhat disorganized, and the information is not clearly arranged, which affects understanding and the effective extraction of key information.\\ 2 points: The report lacks a clear structure, and the sequence of information is chaotic, making it difficult to understand or extract key points.\\ 1 point: The report is completely disorganized, lacking basic logical structure, difficult to understand, and cannot provide effective support for clinical decision-making.\end{tabular} \\ \midrule
\textbf{Clinical Guidance Value (CGV)} &
\begin{tabular}[c]{p{12cm}}5 points: Provides specific and effective treatment recommendations that clearly guide the clinician’s treatment plan and follow-up strategy, offering high clinical value.\\ 4 points: Provides some treatment recommendations, though they may be somewhat general or lacking in detail, still offering useful support for clinical decisions.\\ 3 points: Treatment recommendations are not sufficiently clear or lack key details, potentially impacting the treatment plan, but still provide some guidance.\\ 2 points: Treatment recommendations are overly simplistic or unclear, lacking specificity, and may impact clinical decision-making.\\ 1 point: The report severely lacks critical information and cannot effectively support clinical decisions.\end{tabular} \\ 
\bottomrule
\end{tabular}
\end{table*}
\begin{enumerate}
    \item \textbf{Diagnostic Accuracy} evaluates the correctness of the diagnostic conclusions in the report, particularly the ability to accurately identify the type, location, and severity of rib fractures. A high-quality report should provide a detailed description of the fracture type, location, and healing status, such as: ``Non-displaced fracture of the left 5th rib anterior segment, with bone callus growth, no pneumothorax or hemothorax.'' A low-quality report may omit key information or make incorrect diagnoses, such as: ``Left rib fracture,'' without providing specific location, type, or severity, which can affect treatment decisions. Accurate diagnostic conclusions are critical for clinical decision-making, making diagnostic accuracy a core metric for report quality.
    \item \textbf{Content Completeness} evaluates whether the report covers all necessary medical information, including image findings, diagnostic conclusions, relevant medical knowledge, and treatment recommendations. A high-quality report should comprehensively describe imaging features, diagnostic results, and treatment plans, for example: ``CT shows a fracture of the left 5th rib anterior segment, with bone callus growth, recommended re-examination in 4 weeks.'' A low-quality report may omit crucial information, such as merely stating ``left rib fracture,'' without detailing fracture type, location, or severity, which affects clinical judgment and treatment decisions. Therefore, the completeness of the report content is essential for ensuring accurate diagnosis and treatment.
    \item \textbf{Logical Coherence and Consistency} evaluates whether the report has a rational structure, clear hierarchical organization, and coherent expression. A high-quality report should follow a clear logical sequence, such as imaging findings $\rightarrow$ diagnostic conclusions $\rightarrow$ clinical recommendations, ensuring smooth content flow that is easy to understand. For example: ``CT shows a fracture of the left 5th rib anterior segment, with bone callus growth, recommended re-examination in 4 weeks.'' A low-quality report may suffer from confusion or lack of structure, making it difficult for readers to quickly extract key information, such as: ``Fracture, rest recommended. CT shows left 5th rib fracture.'' Logical coherence and consistency ensure that the report is easy for clinicians to interpret, enabling accurate clinical decision-making.
    \item \textbf{Clinical Guidance Value} evaluates whether the report provides clear and actionable treatment recommendations to help formulate appropriate treatment plans and follow-up schedules. A high-quality report should provide specific treatment suggestions, such as: ``It is recommended that the patient limit thoracic activity, use NSAIDs for pain relief, and re-evaluate the fracture healing status with CT in 4 weeks.'' A low-quality report may lack detailed treatment plans or provide vague suggestions, such as: ``Patient has a fracture, recommended rest, regular follow-up.'' This metric focuses on whether the report has practical clinical value, helping physicians make informed clinical decisions.
\end{enumerate}
Based on these metrics, we have established a scoring system with a range from 1 to 5 (as detailed in Table \ref{tabel1}), representing evaluation levels from low to high, to ensure the independence, objectivity, and scientific rigor of the evaluation process. These metrics ensure that rib fracture diagnostic reports are not only accurate but also clinically practical, helping to improve diagnostic accuracy and providing reliable references for clinical treatment.

\section{Dataset Construction}

\subsection{Data source}

The data used in this study are primarily sourced from three main aspects:
\begin{enumerate}
    \item \textbf{Dataset:} The dataset consists of a publicly accessible rib-fracture CT image dataset used for academic research, obtained from a third-party online repository. It contains approximately \textbf{28.7k} CT images. All images underwent standardized preprocessing, including resizing to \textbf{512$\times$512} pixels and contrast enhancement using windowing techniques. The dataset provides \textbf{YOLO-format} detection annotations (bounding boxes and subtype labels) in normalized (relative) coordinates. These annotations cover five types of rib fractures---\textit{displaced}, \textit{non-displaced}, \textit{buckle}, \textit{segmental}, and \textit{uncertain-type} rib fractures---enabling accurate localization and classification of fracture regions for the detection task. As a precaution, we screened the images for potential patient-identifying information embedded in the image content (e.g., burned-in text). If any such identifiers were detected, the corresponding samples were removed and/or masked prior to model training and evaluation. We did not access, collect, or attempt to infer personally identifiable information, and results are reported only in aggregate form.
    
    \item \textbf{Expert Diagnosis Report Dataset:} To provide high-quality text supervision, we \textbf{invited} five radiology experts from \textbf{West China Hospital (Huaxi)} to perform a second-round review of the CT images and to write corresponding \textbf{preliminary diagnosis reports}. All reports were standardized into a uniform template and text-cleaned, establishing a one-to-one correspondence with the CT images. The reports summarize the experts’ initial judgments and clinical experience regarding fracture-related imaging features, serving as professional references for subsequent diagnostic report generation and evaluation. \textit{Note that these expert reports were created for this study and are not part of the original third-party image repository.}
    
    \item \textbf{Rib Orthopedic Knowledge Base:} The rib orthopedic knowledge base was \textbf{curated from authoritative public sources} (e.g., \emph{Surgery} and the \emph{Rib Fracture Diagnosis and Treatment Guidelines}) and \textbf{reviewed} by clinical experts from \textbf{West China Hospital (Huaxi)}. It systematically organizes multidimensional medical knowledge on fracture classification, etiology, injury mechanisms, clinical manifestations, treatment strategies, and complication management. The knowledge base is stored in a structured format (knowledge graph) to provide authoritative support for knowledge retrieval in the multimodal diagnostic report generation process.
\end{enumerate}

\subsection{Data Augmentation}
\begin{figure*}
    \centering
    \includegraphics[width=1\linewidth]{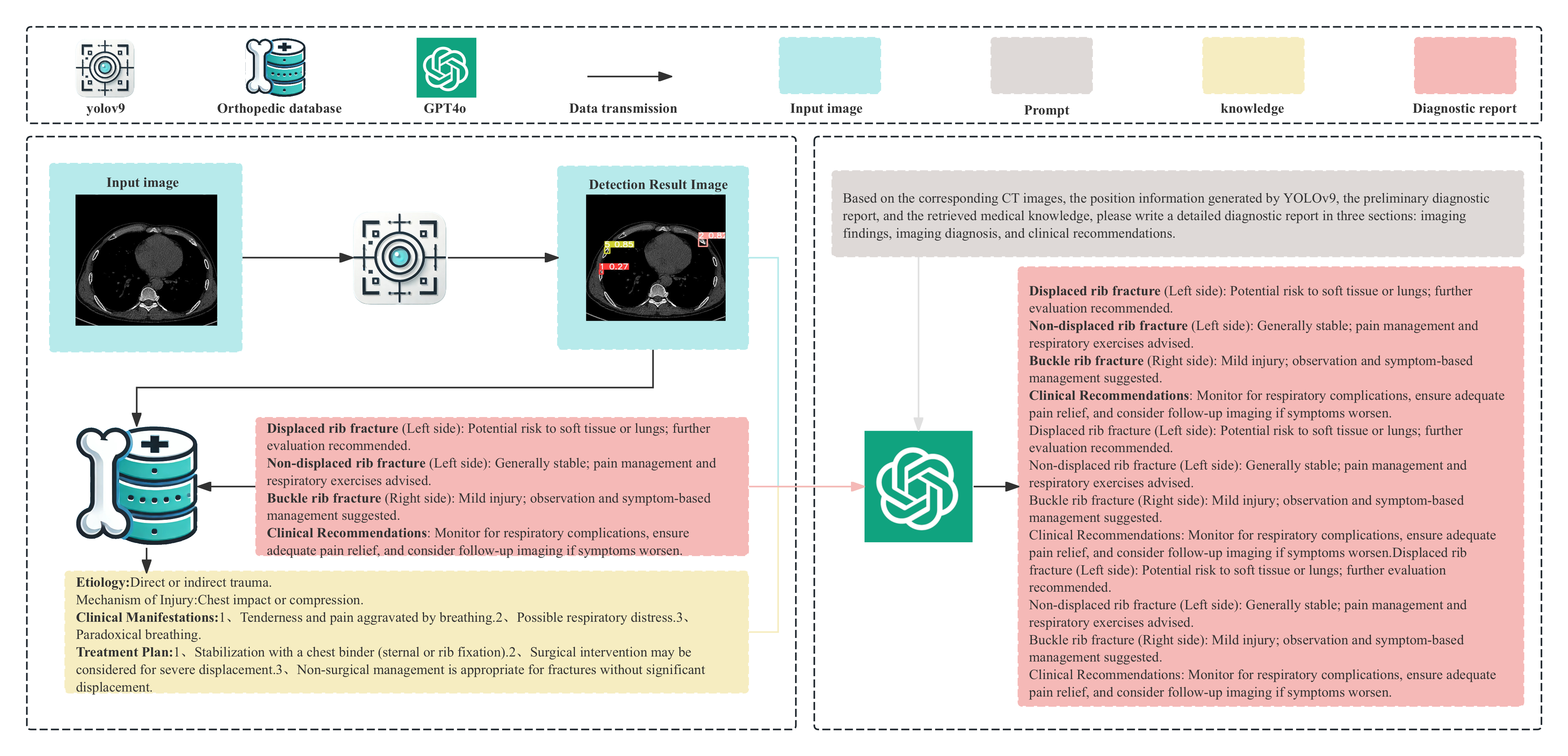}
    \caption{Data construction diagram}
    \label{fig:2}
\end{figure*}
Despite the availability of YOLO-based fracture labels and preliminary diagnostic reports, these data alone are insufficient to support the training of a multimodal large model for rib fracture diagnosis. To generate more comprehensive and detailed diagnostic reports, we have adopted successful practices from the healthcare and education fields, utilizing GPT technology to enhance the diagnostic reports. The specific method is outlined as follows:

First, the CT images are input into the YOLOv9 model to extract fracture location information and type labels. These results, combined with the existing preliminary diagnostic reports, are used to perform relevant knowledge retrieval from the rib orthopedic knowledge base (the specific retrieval method is described in Section 5.2). Next, we integrate the CT images, the location information generated by YOLOv9, the preliminary diagnostic reports, and the retrieved medical knowledge into a complete prompt. This prompt is then input into the GPT-4o model to generate a more detailed and clinically relevant diagnostic report. This process provides more comprehensive training samples for subsequent model training (see Figure \ref{fig:2}).

\subsection{Dataset Evaluation}

Based on the study by Wang et al., this research adopts a large language model (LLM) as an automatic scoring tool to reduce reliance on manual evaluation. Specifically, we randomly selected 1,000 samples from the dataset and used GPT-4o to evaluate them according to predefined scoring criteria. Additionally, five orthopedic experts with extensive clinical experience independently evaluated 200 samples and assessed various metrics. Given the wide range of scoring criteria and the potential variability among evaluators, we used Cohen’s Kappa coefficient \cite{landis1977measurement44} to measure the consistency of scores. The Kappa coefficient ranges from -1 (completely disagree) to 1 (completely agree), with 0 indicating agreement by chance. The experimental results show that the Kappa value between GPT-4o-generated scores and expert scores was 0.74, while the Kappa value between different experts was 0.82. This indicates a high level of consistency between evaluators and between the model and expert evaluations, validating the reliability of the evaluation method (detailed scores are shown in Table \ref{tab:2}). Furthermore, the average expert score was 4.23, and the average GPT-4o score was 4.31, further demonstrating that the enhanced diagnostic reports are of high quality and can serve as effective training samples for the model.

\begin{table}[h!]
\centering
\begin{tabular}{cccccc}
\toprule
\textbf{Model} & \textbf{DA} & \textbf{CC} & \textbf{LCC} & \textbf{CGV} \\ \midrule
GPT4o scorer & 3.67 & 3.92 & 4.01 & 3.94 \\
Expert rating & 3.72 & 3.94 & 3.89 & 4.04 \\ \bottomrule
\end{tabular}
\caption{Dataset Scoring}
\label{tab:2}
\end{table}
\section{Experimental Methodology}
\subsection{Rib Fracture Detection Model Training}
To accurately detect the location and type of rib fractures in CT images, this study employs the YOLOv9 model for object detection. We used an annotated dataset that includes various types of rib fractures, such as displaced and non-displaced fractures. Each CT image in the dataset is precisely annotated, providing information on the fracture type and location. To optimize the model’s detection performance, all input CT images are first preprocessed, which includes resizing the images to 512x512 pixels and applying contrast enhancement techniques to highlight the fracture regions. This preprocessing step helps YOLOv9 effectively recognize and localize the fractures.

During the training process, YOLOv9 extracts features from the CT images and learns to identify the specific location and type of fractures. To ensure detection accuracy, we used cross-entropy loss as the loss function and applied the Adam optimizer to adjust the model’s parameters. The training process aims to improve the model’s fracture detection capabilities, enabling it to handle a variety of fracture types and locations.
\begin{figure*}
    \centering
    \includegraphics[width=1\linewidth]{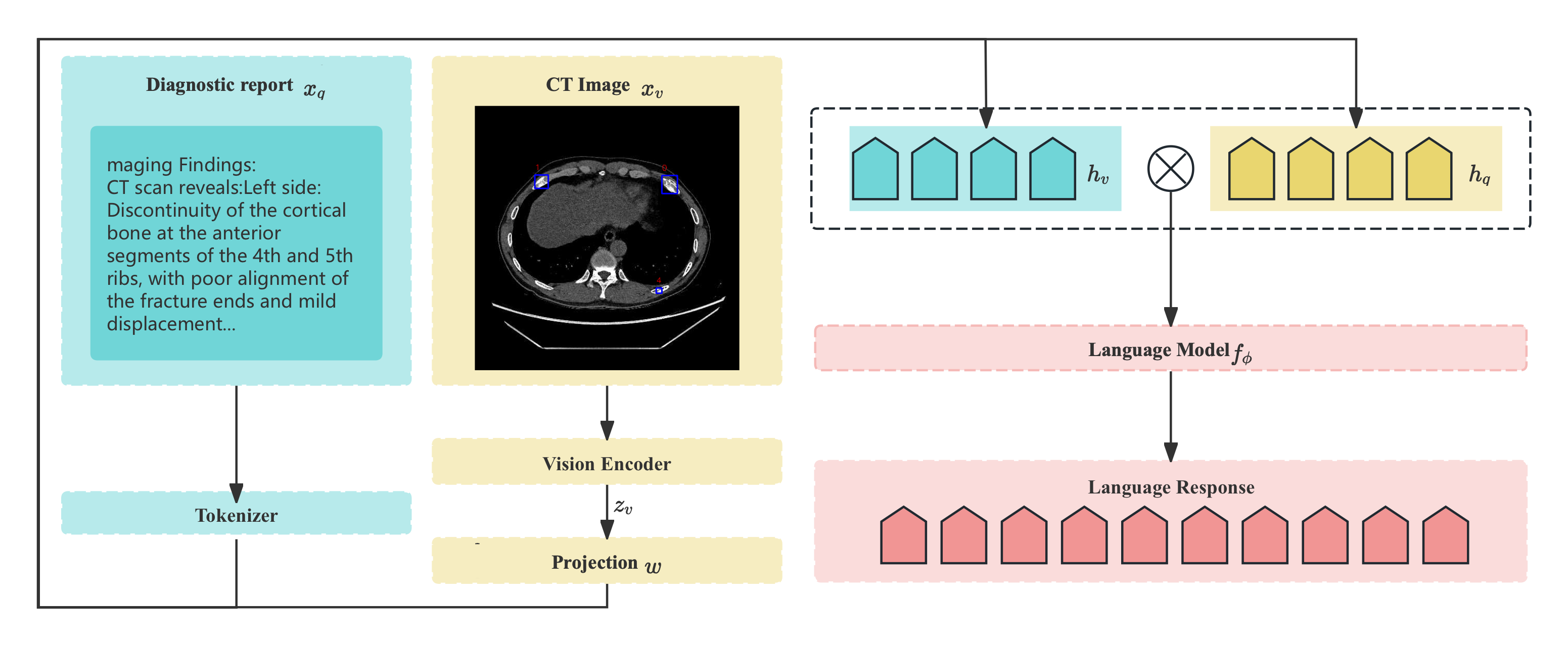}
    \caption{Model architecture diagram}
    \label{fig:3}
\end{figure*}
Through these training steps, the YOLOv9 model can efficiently extract fracture location information and type labels from new CT images. Once trained, the model provides accurate fracture localization and classification, which serves as key information for the subsequent diagnostic report generation.
\subsection{Knowledge Retrieval Methodology}
This section describes a knowledge retrieval and integration method based on the orthopedic knowledge base to enhance the diagnostic capability for rib fractures. The method consists of the following key steps: fracture type extraction and sub-knowledge base selection, query text construction, query vector encoding, knowledge node embedding construction, and similarity-based knowledge retrieval.

\textbf{Fracture Type Extraction and Sub-Knowledge Base Selection}: Based on the YOLO detection results, let \( T \) represent the set of detected rib fracture types:
\begin{equation}
T = \{t_1, t_2, \dots, t_m\}
\end{equation}
For example, if the detected fracture types include "displaced rib fractures" and "segmental rib fractures", then:
\begin{equation}
T = \{\text{displaced rib fractures}, \text{segmental rib fractures}\}
\end{equation}
For each fracture type \( t \in T \), we select the corresponding sub-knowledge base \( K_t \), which contains orthopedic information related to the fracture type, such as classification, etiology, injury mechanism, clinical manifestations, treatment plans, and complication management.

\textbf{Query Text Construction}: Let \( x_E \) represent the preliminary diagnostic report text written by experts. Simultaneously, using the mapping function \( \text{map}_t(\cdot) \), we extract fracture-related descriptive information \( x_{D,t} \) from the detection results. The final query text \( x_{Q,t} \) is constructed as:
\begin{equation}
x_{Q,t} = x_E \oplus x_{D,t}
\end{equation}
where \( \oplus \) represents the text concatenation operation.

\textbf{Query Vector Encoding}: A pre-trained orthopedic text encoder \( f(\cdot) \) (such as a BERT encoder fine-tuned for the orthopedic domain) is used to map the query text to the vector space, resulting in the query vector:
\begin{equation}
q_t = f(x_{Q,t}) \in \mathbb{R}^d
\end{equation}
where \( d \) is the dimension of the vector embedding.

\textbf{Knowledge Node Embedding Construction}: In the sub-knowledge base \( K_t \), each orthopedic knowledge node \( i \) contains structured orthopedic information related to the fracture type. Its text representation \( x_i \) is constructed as:
\begin{equation}
\begin{aligned}
x_i = &\; x_{\text{classification},i} \oplus x_{\text{etiology},i} \\
      &\oplus x_{\text{injury mechanism},i} \oplus x_{\text{clinical manifestations},i} \\
      &\oplus x_{\text{treatment plan},i} \oplus x_{\text{complication management},i}
\end{aligned}
\end{equation}

Then, using the same text encoder \( f(\cdot) \), the text representation is converted into an embedding vector:
\begin{equation}
v_i(t) = f(x_i) \in \mathbb{R}^d
\end{equation}

\textbf{Similarity-Based Knowledge Retrieval}: For each detected fracture type \( t \), the query vector \( q_t \) and the knowledge node embedding vector \( v_i(t) \) are compared using cosine similarity:
\begin{equation}
\text{sim}(q_t, v_i(t)) = \frac{q_t \cdot v_i(t)}{\|q_t\| \|v_i(t)\|}
\end{equation}
Then, in the sub-knowledge base \( K_t \), the top \( k \) knowledge nodes with the highest similarity are selected to form the candidate set:
\begin{equation}
C_t = \{i \mid \text{sim}(q_t, v_i(t)) \text{ ranks in the top } k\}
\end{equation}
where \( C_t \) represents the candidate knowledge node set for fracture type \( t \).

\textbf{Final Knowledge Retrieval Output}: For all detected fracture types \( t \in T \), this method retrieves the corresponding candidate knowledge sets \( C_t \), which provide important information to support the multimodal diagnostic report generation.

By integrating fracture detection results, expert-written reports, and an orthopedic knowledge base, this method not only improves the accuracy of diagnostic reports but also ensures that the model output aligns with authoritative knowledge in the orthopedic field, enhancing the interpretability and clinical reliability of the diagnostic results.
\begin{figure*}
    \centering
    \includegraphics[width=1\linewidth]{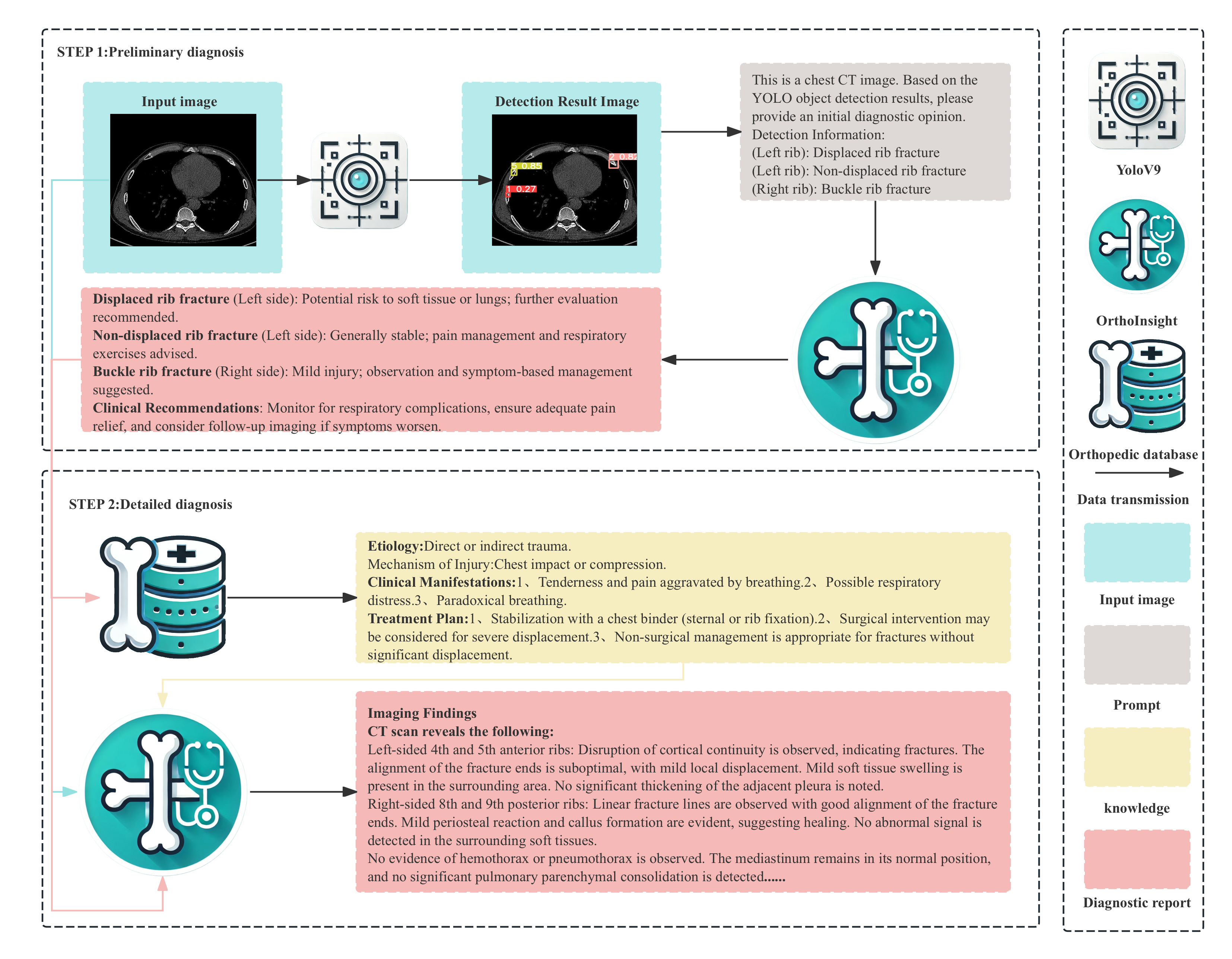}
    \caption{Diagnostic flow chart}
    \label{fig:4}
\end{figure*}
\subsection{Model Architecture and Training Stages}
\subsubsection{Model Architecture}
We adopted LLaVA, a multimodal dialogue model in the general domain, as the initial general language model (LM) for this study. The model architecture is shown in Figure \ref{fig:3} and architecture consists of three main components: a pre-trained vision encoder (ViT), a lightweight projection module (Proj), and a pre-trained large language model (LLM). Specifically, the CT image \( X_v \) and the diagnostic report text \( x_q \) are processed as follows: The CT image is first passed through the vision encoder to obtain visual feature representations \( Z_v \), which are then mapped into visual embeddings \( h_v \) using the projection matrix \( W \). Meanwhile, the diagnostic report \( x_q \) is tokenized by the Tokenizer, resulting in the text feature representation \( h_q \). Next, the visual embeddings \( h_v \) and text embeddings \( h_q \) interact, typically through a dot product operation (\( \odot \)), to generate a joint representation. This joint representation is then passed through the pre-trained language model \( f_\phi \) to generate the final diagnostic report \( y \).

This process can be expressed by the following formula:
\[
\text{Language Response} = f_\phi \left( W \cdot \text{Encoder}(X_v) \odot \text{Tokenizer}(x_q) \right)
\]

\subsubsection{Model Training Phases}
The training process is divided into two stages:
\begin{enumerate}
    \item \textbf{Stage 1: Orthopedic Concept Feature Alignment}: In this stage, we freeze the weights of the vision encoder and the language model and only update the projection matrix. The goal is to align the visual embeddings with the word embeddings in the pre-trained large language model (LLM). This allows us to align a large number of orthopedic-related visual concepts with the text embeddings in the pre-trained language model, thereby enhancing the model’s understanding of the orthopedic domain.
    \item \textbf{Stage 2: Instruction Fine-Tuning}: In the second stage, we freeze the vision encoder weights and continue to update the projection layer and the pre-trained weights of the LLM. The model is fine-tuned using augmented CT image-diagnostic report instruction-following data. This stage employs an end-to-end training approach to further optimize the model’s performance.
\end{enumerate}

\subsection{Inference Phase for Rib Fracture Diagnosis}
After the training phase, the system enters the inference stage, where the trained model is applied to real CT images to generate diagnostic reports for rib fractures. The specific flowchart is shown in Figure \ref{fig:4}. The inference process consists of the following key steps:
\begin{enumerate}
    \item The CT images to be analyzed are input into the YOLOv9 model for fracture detection. The YOLOv9 model outputs fracture location information (i.e., the bounding boxes of the fractures) and fracture types (such as displaced fractures, non-displaced fractures, etc.).
    \item The fracture location information and fracture type obtained from YOLOv9 are input into the orthopedic knowledge base for relevant knowledge retrieval. The orthopedic knowledge base contains related orthopedic information for different types of fractures, such as fracture causes, injury mechanisms, treatment plans, and complication management.
    \item The CT image fracture location information, fracture type, and the orthopedic knowledge retrieved from the knowledge base are input into the trained LLaVA 1.5-7B model. The model combines image features, textual descriptions, and orthopedic knowledge to generate a detailed diagnostic report. The report includes a detailed description of the fracture, treatment recommendations, possible complications, and follow-up plans.
\end{enumerate}

\section{Experimental Setups}
\subsection{Datasets}
We evaluate all models on a publicly accessible rib-fracture CT image dataset obtained from a third-party online repository. It comprises 28{,}675 chest CT images at $512\times512$ resolution, each annotated in YOLO format with five fracture subtypes---displaced, non-displaced, buckle, segmental, and indeterminate. All samples are divided into training, test, and validation sets with a ratio of 7:2:1. To provide textual supervision for diagnostic report generation, we invited five radiology experts from West China Hospital (Huaxi) to write preliminary diagnosis reports based on the CT images (and corresponding detection annotations when applicable). The expert reports were standardized into a uniform template and text-cleaned, establishing a one-to-one correspondence with the CT images; note that these expert reports were created specifically for this study and are not part of the original third-party repository. In addition, an orthopaedic knowledge graph containing six node types and about 42k triples is curated from authoritative public sources (e.g., \emph{Surgery} and the \emph{Guideline for Rib-Fracture Management}) and reviewed by clinical experts to support medical fact retrieval. During preprocessing, YOLOv9 first localises and classifies fracture sites; its outputs, together with the expert reports and retrieved graph facts, are merged via a templated prompt and expanded (using GPT-4o) into structured, fine-grained diagnostic reports, which serve as training targets for the report-generation task.

\subsection{Evaluation Metrics}

\paragraph{Detection.}
For the rib-fracture detection task, we follow the standard object-detection protocol and report four metrics:  
\textbf{Precision}~(P), \textbf{Recall}~(R), \textbf{mAP@0.5} (mean Average Precision at an IoU threshold of~0.5), and \textbf{mAP@0.5:0.95} (mAP averaged over IoU thresholds 0.50–0.95 in steps of~0.05).  
The metric mAP@0.5 measures overall detection accuracy when $\mathrm{IoU}\!\ge\!0.50$, whereas mAP@0.5:0.95 is more stringent and reflects the model’s sensitivity to localisation errors.  
All scores are computed on the test split using the official COCO/YOLO evaluation script.

\paragraph{Report generation.}
Quality of the generated diagnostic reports is assessed with a four-dimensional rubric recommended by orthopaedic specialists, comprising \textbf{Diagnostic Accuracy (DA)}—correctness of fracture type and anatomical location; \textbf{Content Completeness (CC)}—inclusion of image findings, diagnostic conclusion, and follow-up advice; \textbf{Logical Coherence \& Consistency (LCC)}—structural clarity and narrative fluency; and \textbf{Clinical Guidance Value (CGV)}—practical utility of the recommended treatment plan. Each criterion is rated on a five-point Likert scale (1\,=\,poor, 5\,=\,excellent). A random sample of 1\,000 model outputs is automatically scored with GPT-4o, and five attending orthopaedists independently evaluate 200 of them. Agreement between GPT-4o and the human panel yields Cohen’s $\kappa = 0.74$, while inter-expert agreement reaches $\kappa = 0.82$, indicating good consistency of the evaluation protocol.

\subsection{Baseline}
To comprehensively evaluate the performance of OrthoInsight, we selected six mainstream baseline models for comparative analysis, including GPT-4 \cite{openai2024gpt4}, Claude-3 \cite{anthropic2024claude3}, LLAVA1.5-7B \cite{liu2024improvedbaselinesvisualinstruction}, VisualGLM-6b \cite{du2022glm} \cite{ding2021cogview}, Qwen2-VL-7B \cite{wang2024qwen2vlenhancingvisionlanguagemodels}, and MiniGPT-4 \cite{zhu2023minigpt4enhancingvisionlanguageunderstanding}. These models represent the cutting-edge technologies in current multimodal and language generation tasks, encompassing different model architectures and optimization strategies, thus providing a multidimensional reference for the comparative analysis.

\subsection{Implementation Details}
In our experiments, we conduct all training and inference on a workstation running Ubuntu~20.04, equipped with an Intel(R) Xeon(R) Platinum 8255C CPU (12 vCPUs @ 2.50GHz) and three NVIDIA RTX~3090 GPUs (24,GB). All models are implemented in Python 3.8 with PyTorch 1.11.0 and deployed using CUDA 11.3. The dataset is partitioned into training, validation, and test splits with a ratio of 7:2:1. Although images are standardized to $512\times512$ during dataset preprocessing, YOLOv9 is trained with $640\times640$ input using letterbox resizing/padding to fit the network input. For the object detection task, we adopt YOLOv9 as the baseline detector. The model is trained with an input image size of $640 \times 640$, a batch size of 16, and 150 training epochs. The initial learning rate is set to 0.01 using cosine annealing. For the report generation task, we use LLaVA-1.5-7B as the base multimodal model and apply parameter-efficient fine-tuning via LoRA. The LoRA configuration includes a rank of 64, an $\alpha$ of 16, and a dropout rate of 0.05. We train for 20 epochs with a batch size of 10 and a learning rate of $1\mathrm{e}{-4}$, requiring approximately 60 hours on three RTX~3090 GPUs.

\section{Experiments}
To comprehensively validate the effectiveness of the proposed OrthoInsight framework, we design a series of experiments centered around three key evaluation tasks, leading to the following research questions:

\textbf{RQ1:} Can the YOLOv9 model adopted in OrthoInsight achieve high-precision localization and classification in rib fracture detection?

\textbf{RQ2:} In diagnostic report generation, does OrthoInsight significantly outperform existing multimodal baselines in terms of content accuracy, completeness, and clinical applicability?

\textbf{RQ3:} Do the core components of OrthoInsight—such as knowledge graph retrieval, expert report integration, and multi-stage instruction design—contribute substantially to the final report quality?
\subsection{Experimental Validation of YOLOv9 Model(RQ1)}
To validate the reliability of the upstream detection component in our pipeline, we adopt YOLOv9 as the rib-fracture detector and evaluate its performance on the detection task. In this paper, RQ1 reports the detection accuracy of the adopted YOLOv9 module as a prerequisite for providing fracture localization and subtype cues to downstream diagnostic report generation, rather than proposing a new detection algorithm. In our experiments, YOLOv9 achieves a precision (P) of 98.3\%, a recall (R) of 89\%, and a mean average precision (mAP) at an IoU threshold of 0.5 of 97.2\%. These results indicate that the detector can provide effective fracture localization and classification cues for subsequent multimodal report generation.9

\subsection{Evaluation of Diagnostic Reports RQ2}
We redesigned the scoring process by introducing a large language model (LLM)-based automated scoring mechanism to reduce reliance on manual annotations. Specifically, we randomly selected 1,000 samples from the dataset and used OpenAI’s GPT-4o for automated scoring. To ensure the reliability of the scoring results, five senior orthopedic experts were invited to manually score 200 samples from these based on predefined evaluation criteria. By combining automated scoring and human evaluation, this study maintained scoring efficiency while enhancing the objectivity and consistency of the evaluation results. To further validate the reliability of the evaluation results, three statistical methods—Kendall’s Tau, Pearson correlation coefficient, and Fleiss’ Kappa—were applied to analyze the consistency of expert ratings, with the results shown in Table \ref{tabel3}. The analysis indicated a high level of consistency, with Kendall’s Tau coefficients all above 0.7 (accuracy 0.85, clarity 0.88), Pearson correlation coefficients close to 0.9 (accuracy 0.92, overall rating 0.91), and Fleiss’ Kappa values all above 0.7 (clarity 0.80, accuracy 0.76). 

\begin{table}[htbp]
  \centering
  \caption{Consistency Analysis of Scoring Results}
  \begin{tabular}{lccc}
    \toprule
    \textbf{Metric} & \textbf{Kendall’s Tau} & \textbf{Pearson} & \textbf{Fleiss’ Kappa} \\
    \midrule
    DA & 0.85 & 0.92 & 0.76 \\
    CC & 0.88 & 0.90 & 0.80 \\
    LCC & 0.73 & 0.85 & 0.70 \\
    CGV & 0.78 & 0.87 & 0.72 \\
    Average & 0.82 & 0.91 & 0.74 \\
    \bottomrule
  \end{tabular}
  \label{tabel3}
\end{table}

These findings further validated the reliability and consistency of the scoring process. In terms of specific scores, OrthoInsight achieved scores of 4.32, 4.11, 4.41, and 4.27 for Diagnostic Accuracy, Content Completeness, Logical Coherence and Consistency, and Clinical Guidance Value, respectively, with an average score of 4.28 (as shown in Table \ref{table4}). All evaluation metrics scored above 4, demonstrating the model’s exceptional performance in generating high-quality diagnostic reports. In contrast, although GPT-4 and Claude-3 performed well in some metrics, their overall scores were slightly lower, highlighting the advantages of our knowledge enhancement approach and customized model design. Additionally, LLAVA1.5-7B, as the base model for OrthoInsight, achieved the best performance among models of a similar scale, further demonstrating its strength and reliability as a foundational model.
\begin{table}
\caption{\label{table4} Original Score Sheet}
\centering
\begin{tabular}{@{}lcccccc@{}}
    \hline
    \rotatebox{90}{} & \textbf{Model} & \textbf{DA.} & \textbf{CC.} & \textbf{LCC.} & \textbf{CGV.} & \textbf{Avg.} \\
    \hline
    \multirow{7}{*}{\rotatebox{90}{\textbf{GPT4o scorer}}} & GPT4 & 3.51 & 3.29 & 3.41 & 3.27 & 3.37 \\
    & Claude3 & 3.26 & 3.17 & 3.56 & 3.21 & 3.30 \\
    & LLAVA1.5-7B & 3.02 & 2.97 & 3.16 & 2.83 & 3.05 \\
    & VisualGLM-6B & 2.83 & 3.04 & 2.81 & 2.83 & 2.89 \\
    & Qwen2-VL-7B & 2.79 & 2.71 & 3.03 & 2.79 & 2.83 \\
    & MiniGPT-4 & 2.28 & 2.38 & 2.48 & 2.51 & 2.41 \\
    & \textbf{OrthoInsight}& \textbf{4.41} & \textbf{4.16} & \textbf{4.44} & \textbf{4.39} & \textbf{4.35} \\
    \hline
    \multirow{7}{*}{\rotatebox{90}{\textbf{Expert rating}}} & GPT4           & 3.43 & 3.37 & 3.35 & 3.25 & 3.35 \\
& Claude3        & 3.30 & 3.46 & 3.50 & 3.23 & 3.38 \\
& LLAVA1.5-7B         & 2.94 & 3.01 & 3.11 & 2.87 & 2.98 \\
& VisualGLM-6B       & 2.83 & 3.04 & 2.91 & 2.83 & 2.90 \\
& Qwen2-VL-7B     & 2.79 & 2.76 & 3.03 & 2.86 & 2.86 \\
& MiniGPT-4   & 2.33 & 2.42 & 2.53 & 2.46 & 2.44 \\
& \textbf{OrthoInsight} & \textbf{4.32} & \textbf{4.11} & \textbf{4.41} & \textbf{4.27} & \textbf{4.28} \\
    \hline
\end{tabular}
\end{table}

\begin{table}[htbp]
\caption{Ablation Experiment Score Comparison Table}
\begin{center}
\begin{tabular}{lccccc}
\toprule
\textbf{Model} & \textbf{Acc.} & \textbf{Cla.} & \textbf{Com.} & \textbf{Pra.} & \textbf{Ave.} \\
\midrule
Original & 2.94 & 3.01 & 3.11 & 2.87 & 2.98 \\
SFT & 3.62 & 3.54 & 3.64 & 3.53 & 3.58 \\
COT & 3.07 & 3.09 & 3.22 & 3.01 & 3.10 \\
KE. & 3.23 & 3.27 & 3.24 & 3.26 & 3.25 \\
SFT-COT & 3.71 & 3.62 & 3.74 & 3.66 & 3.68 \\
SFT-KE. & \textbf{4.41} & \textbf{4.16} & \textbf{4.44} & \textbf{4.39} & \textbf{4.35} \\
\bottomrule
\end{tabular}
\label{tab5}
\end{center}
\end{table}

\subsection{Ablation Experiment (RQ3)}
\begin{table}[h]
  \caption{\label{table6} Sampled case}
  \centering
  \sethlcolor{lime}
  \renewcommand{\arraystretch}{1}
  \scriptsize
  \begin{tabular}{p{1cm} p{7cm}}
    \hline
    \textbf{Diagnostic picture} &
      \includegraphics[width=0.3\textwidth]{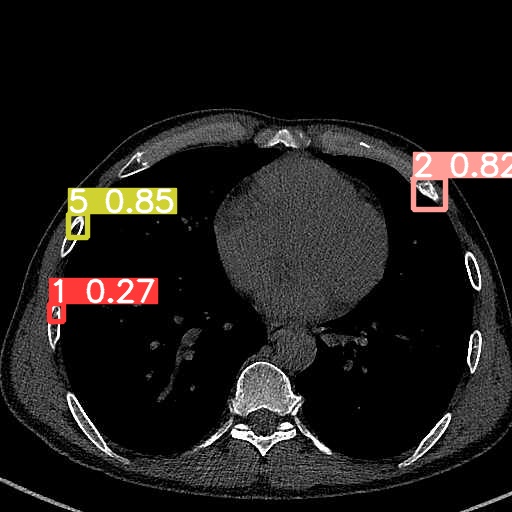}\\
    \hline

    \textbf{Diagnostic report}&
    \textbf{Imaging Findings:}\\
    & 1.~Discontinuity of the cortical bone in the anterior segments of the left 4\textsuperscript{th} and 5\textsuperscript{th} ribs, indicating fractures. The alignment of the fracture ends is sub‑optimal, with mild displacement. Mild soft‑tissue swelling is observed in the surrounding area, while no significant thickening of the adjacent pleura is noted.\\
    & 2.~Linear fracture shadows are observed in the posterior segments of the right 8\textsuperscript{th} and 9\textsuperscript{th} ribs. The fracture ends are well‑aligned, and mild bone hyperplasia with callus formation is visible. No abnormal signal is detected in the surrounding soft tissues.\\
    & 3.~No obvious signs of hemothorax or pneumothorax are observed. The mediastinum remains in its normal position, and no distinct consolidation or abnormal opacities are detected in the lung parenchyma.\\[5pt]\cline{2-2}

    & \textbf{Imaging Diagnosis:}\\
    & 1.~Acute fractures of the anterior segments of the left 4\textsuperscript{th} and 5\textsuperscript{th} ribs, accompanied by mild soft‑tissue swelling. No apparent signs of hemothorax or pneumothorax.\\
    & 2.~Old fractures of the posterior segments of the right 8\textsuperscript{th} and 9\textsuperscript{th} ribs, with callus formation and good healing progression.\\
    & 3.~No mediastinal displacement or pulmonary parenchymal lesions detected.\\[5pt]\cline{2-2}

    & \textbf{Clinical Recommendations:}\\
    & 1.~\textbf{Activity restriction:} Limit thoracic movements and avoid vigorous coughing or significant postural changes to reduce the risk of further displacement of the fracture ends.\\
    & 2.~\textbf{Pain management:} Non‑steroidal anti‑inflammatory drugs (NSAIDs) may be used for analgesia. If necessary, elastic chest bandaging can be considered to alleviate chest‑wall pain and prevent respiratory distress.\\
    & 3.~\textbf{Complication monitoring:} Observe for any signs of delayed pneumothorax or hemothorax, especially if symptoms such as dyspnea, persistent chest pain, or decreased breath sounds develop.\\
    & 4.~\textbf{Imaging follow‑up:} A follow‑up chest CT is recommended in 2–4 weeks to assess fracture healing. If displacement risk persists, further therapeutic interventions may be considered.\\
    \hline
  \end{tabular}
\end{table}
To validate the effectiveness of each method, we designed and conducted ablation experiments, testing SFT only, knowledge enhancement only, and replacing knowledge enhancement with Zero-shot-CoT to analyze the contribution of different modules to model performance. As shown in Table \ref{tab5}, the SFT method demonstrated the most significant improvement across various performance metrics. By applying SFT, the model was better adapted to specific task scenarios, achieving targeted optimization and exhibiting higher accuracy and stability in professional tasks. Although the performance of the knowledge enhancement method alone was not as strong as that of SFT, it still played a positive role in supplementing domain knowledge and enhancing task comprehension. When knowledge enhancement was combined with SFT, the model’s overall performance improved further, validating the effectiveness of their synergy. This indicates that the knowledge enhancement method provides the model with enriched background information and professional prompts, helping to generate more accurate and comprehensive responses. In contrast, the performance of the Zero-shot-CoT method was relatively limited, possibly due to the lack of domain-specific prompts, which hindered its task comprehension and detailed response generation. This further demonstrates the importance of professional knowledge prompts in multimodal tasks and complex professional scenarios. Overall, the experimental results show that both SFT and knowledge enhancement play crucial roles in improving model performance, and their combination offers significant advantages in enhancing model adaptability and overall performance.

\section{Case Study}
This section presents a CT imaging diagnostic report generated by OrthoInsight, with high-quality diagnostic content highlighted in green. Table \ref{table6} provides a detailed case example. In this case, OrthoInsight successfully identified common types of rib fractures and accurately assessed their freshness, alignment status, and risk of complications. Specifically, the system not only differentiates between fresh and old fractures but also provides detailed analyses of fracture stability, local soft tissue response, and potential complications. Compared to traditional reports derived from standard radiology datasets, OrthoInsight’s diagnostic report demonstrates greater professionalism and specificity. Standard reports typically focus on fracture location and basic imaging descriptions. In contrast, OrthoInsight offers a more comprehensive interpretation, including fracture alignment assessment, imaging characteristics of callus formation, soft tissue response, and a complete evaluation of the mediastinum and lungs. Additionally, this report incorporates explicit clinical management recommendations, covering activity restriction, pain management, complication monitoring, and imaging follow-up, thereby enhancing its practicality and applicability.
During the report generation process, OrthoInsight exhibits outstanding imaging analysis capabilities and clinical decision support, aiding radiologists in improving diagnostic efficiency and ensuring that patients receive more targeted treatment plans. This AI-driven imaging reporting system not only enhances the accuracy and depth of imaging diagnostics but also provides clinicians with high-quality reference information, fostering advancements in intelligent medical imaging.

\section{Conclusion}
In this study, we present OrthoInsight, a multi-modal deep learning framework for automated rib fracture diagnosis and reporting from CT images. Integrating YOLOv9 for detection, a medical knowledge graph for contextual understanding, and LLaVA for report generation, OrthoInsight addresses limitations of manual interpretation, such as inefficiency and error-proneness. Evaluated on a comprehensive dataset using rigorous metrics, the framework outperforms state-of-the-art models with an average score of 4.28 across diagnostic accuracy, content completeness, logical coherence, and clinical guidance. OrthoInsight not only accurately detects and classifies rib fractures but also generates detailed, clinically meaningful reports, supporting radiologists in decision-making and improving diagnostic efficiency. While promising, the current implementation is research-oriented and not a substitute for professional medical judgment. Future work will focus on expanding dataset diversity, enhancing model robustness, and exploring broader applications in medical imaging.

\section{Ethics Statement}
This study is conducted solely for research and technical validation. The generated diagnoses and reports do not constitute medical advice and are not intended to replace professional clinical judgment. The CT images used in this work are de-identified; no patient-identifiable information was accessed, collected, or inferred. The expert diagnostic reports used for evaluation were newly created by invited clinicians based on the images for research purposes, rather than being extracted from the original repository. Any potential clinical deployment should follow applicable regulations and institutional policies, and all generated outputs must be reviewed by qualified clinicians before use in real-world practice.
\section{Data source and availability}
The CT images used in this study were obtained from a publicly accessible online repository (accessed on 2025-01-19). During peer review, the de-identified data and corresponding annotations used in our experiments can be provided to the editor and reviewers upon reasonable request under a confidentiality agreement, subject to the terms of the original data source. Upon acceptance, we will release the code, scripts, prompt templates, knowledge-graph construction pipeline, and evaluation toolkit needed to reproduce the reported results.
\bibliographystyle{unsrt}
\bibliography{citations}

@article{landis1977measurement44,
  title={The measurement of observer agreement for categorical data},
  author={Landis JRKoch, GG},
  journal={Biometrics},
  volume={33},
  number={1},
  pages={159174},
  year={1977}
}

@misc{zhu2023minigpt4enhancingvisionlanguageunderstanding,
      title={MiniGPT-4: Enhancing Vision-Language Understanding with Advanced Large Language Models}, 
      author={Deyao Zhu and Jun Chen and Xiaoqian Shen and Xiang Li and Mohamed Elhoseiny},
      year={2023},
      eprint={2304.10592},
      archivePrefix={arXiv},
      primaryClass={cs.CV},
      url={https://arxiv.org/abs/2304.10592}, 
}

@misc{wang2024qwen2vlenhancingvisionlanguagemodels,
      title={Qwen2-VL: Enhancing Vision-Language Model's Perception of the World at Any Resolution}, 
      author={Peng Wang and Shuai Bai and Sinan Tan and Shijie Wang and Zhihao Fan and Jinze Bai},
      year={2024},
      eprint={2409.12191},
      archivePrefix={arXiv},
      primaryClass={cs.CV},
      url={https://arxiv.org/abs/2409.12191}, 
}

@inproceedings{du2022glm,
  title={GLM: General Language Model Pretraining with Autoregressive Blank Infilling},
  author={Du, Zhengxiao and Qian, Yujie and Liu, Xiao and Ding, Ming and Qiu, Jiezhong and Yang, Zhilin and Tang, Jie},
  booktitle={Proceedings of the 60th Annual Meeting of the Association for Computational Linguistics (Volume 1: Long Papers)},
  pages={320--335},
  year={2022}
}

@article{ding2021cogview,
  title={Cogview: Mastering text-to-image generation via transformers},
  author={Ding, Ming and Yang, Zhuoyi and Hong, Wenyi and Zheng, Wendi and Zhou, Chang and Yin, Da and Lin, Junyang and Zou, Xu and Shao, Zhou and Yang, Hongxia and others},
  journal={Advances in Neural Information Processing Systems},
  volume={34},
  pages={19822--19835},
  year={2021}
}

@misc{liu2024improvedbaselinesvisualinstruction,
      title={Improved Baselines with Visual Instruction Tuning}, 
      author={Haotian Liu and Chunyuan Li and Yuheng Li and Yong Jae Lee},
      year={2024},
      eprint={2310.03744},
      archivePrefix={arXiv},
      primaryClass={cs.CV},
      url={https://arxiv.org/abs/2310.03744}, 
}

@misc{anthropic2024claude3,
  author       = {Anthropic},
  title        = {The Claude 3 Model Family: Opus, Sonnet, Haiku},
  year         = {2024},
  url          = {https://www-cdn.anthropic.com/de8ba9b01c9ab7cbabf5c33b80b7bbc618857627/Model_Card_Claude_3.pdf}
}

@misc{zhang2023huatuogpttaminglanguagemodel,
      title={HuatuoGPT, towards Taming Language Model to Be a Doctor}, 
      author={Hongbo Zhang and Junying Chen and Feng Jiang and Fei Yu and Zhihong Chen and Jianquan Li and Guiming Chen and Xiangbo Wu and Zhiyi Zhang and Qingying Xiao and Xiang Wan and Benyou Wang and Haizhou Li},
      year={2023},
      eprint={2305.15075},
      archivePrefix={arXiv},
      primaryClass={cs.CL},
      url={https://arxiv.org/abs/2305.15075}, 
}

@article{karunaratne2022review,
  title={A review of comprehensiveness, user-friendliness, and contribution for sustainable design of whole building environmental life cycle assessment software tools},
  author={Karunaratne, Shiromi and Dharmarathna, Dilshi},
  journal={Building and Environment},
  volume={212},
  pages={108784},
  year={2022},
  publisher={Elsevier}
}

@inproceedings{ghafourian2023readability,
  title={Readability Measures as Predictors of Understandability and Engagement in Searching to Learn},
  author={Ghafourian, Yasin and Hanbury, Allan and Knoth, Petr},
  booktitle={International Conference on Theory and Practice of Digital Libraries},
  pages={173--181},
  year={2023},
  organization={Springer}
}

@article{tiffin2011evaluating,
  title={Evaluating professionalism in medical undergraduates using selected response questions: findings from an item response modelling study},
  author={Tiffin, Paul A and Finn, Gabrielle M and McLachlan, John C},
  journal={BMC medical education},
  volume={11},
  pages={1--9},
  year={2011},
  publisher={Springer}
}

@article{Yang_2025,
   title={Application of large language models in disease diagnosis and treatment},
   journal={Chinese Medical Journal},
   volume={138},
   number={2},
   pages={130--142},
   year={2025},
   doi={10.1097/CM9.0000000000003456},
   author={Yang, Xintian and et al.}
}

@article{Chang_2024,
   title={Use of SNOMED CT in Large Language Models: Scoping Review},
   journal={JMIR Medical Informatics},
   volume={12},
   pages={e62924},
   year={2024},
   month={Oct},
   doi={10.2196/62924},
   author={Chang, Eunsuk and Sung, Sumi}
}

@misc{nguyen2024logramedlongcontextmultigraph,
      title={LoGra-Med: Long Context Multi-Graph Alignment for Medical Vision-Language Model}, 
      author={Duy M. H. Nguyen and Nghiem T. Diep and Trung Q. Nguyen and Hoang-Bao Le and Tai Nguyen and Tien Nguyen and TrungTin Nguyen and Nhat Ho and Pengtao Xie and Roger Wattenhofer and James Zhou and Daniel Sonntag and Mathias Niepert},
      year={2024},
      eprint={2410.02615},
      archivePrefix={arXiv},
      primaryClass={cs.LG},
      url={https://arxiv.org/abs/2410.02615}, 
}

@misc{yang2023dawnlmmspreliminaryexplorations,
      title={The Dawn of LMMs: Preliminary Explorations with GPT-4V(ision)}, 
      author={Zhengyuan Yang and Linjie Li and Kevin Lin and Jianfeng Wang and Chung-Ching Lin and Zicheng Liu and Lijuan Wang},
      year={2023},
      eprint={2309.17421},
      archivePrefix={arXiv},
      primaryClass={cs.CV},
      url={https://arxiv.org/abs/2309.17421}, 
}

@misc{he2025pefomedparameterefficientfinetuning,
      title={PeFoMed: Parameter Efficient Fine-tuning of Multimodal Large Language Models for Medical Imaging}, 
      author={Jinlong He and Pengfei Li and Gang Liu and Genrong He and Zhaolin Chen and Shenjun Zhong},
      year={2025},
      eprint={2401.02797},
      archivePrefix={arXiv},
      primaryClass={cs.CL},
      url={https://arxiv.org/abs/2401.02797}, 
}

@misc{li2023llavamedtraininglargelanguageandvision,
      title={LLaVA-Med: Training a Large Language-and-Vision Assistant for Biomedicine in One Day}, 
      author={Chunyuan Li and Cliff Wong and Sheng Zhang and Naoto Usuyama and Haotian Liu and Jianwei Yang and Tristan Naumann and Hoifung Poon and Jianfeng Gao},
      year={2023},
      eprint={2306.00890},
      archivePrefix={arXiv},
      primaryClass={cs.CV},
      url={https://arxiv.org/abs/2306.00890}, 
}

@misc{chen2024advancinghighresolutionvisionlanguage,
      title={Advancing High Resolution Vision-Language Models in Biomedicine}, 
      author={Zekai Chen and Arda Pekis and Kevin Brown},
      year={2024},
      eprint={2406.09454},
      archivePrefix={arXiv},
      primaryClass={cs.CL},
      url={https://arxiv.org/abs/2406.09454}, 
}

@misc{han2025medalpacaopensourcecollection,
      title={MedAlpaca -- An Open-Source Collection of Medical Conversational AI Models and Training Data}, 
      author={Tianyu Han and Lisa C. Adams and Jens-Michalis Papaioannou and Paul Grundmann and Tom Oberhauser and Alexei Figueroa and Alexander Löser and Daniel Truhn and Keno K. Bressem},
      year={2025},
      eprint={2304.08247},
      archivePrefix={arXiv},
      primaryClass={cs.CL},
      url={https://arxiv.org/abs/2304.08247}, 
}

@article{Yang_2024,
   title={Unmasking and quantifying racial bias of large language models in medical report generation},
   journal={Commun Med},
   volume={4},
   pages={176},
   year={2024},
   doi={https://doi.org/10.1038/s43856-024-00601-z},
   author={Yang, Y. and Liu, X. and Jin, Q. and et al.}
}

@article{Hasan_2020,
   title={Knowledge Graph-Enabled Cancer Data Analytics},
   journal={IEEE Journal of Biomedical and Health Informatics},
   volume={24},
   number={7},
   pages={1952--1967},
   year={2020},
   doi={10.1109/JBHI.2020.2990797},
   author={Hasan, S.M. Shamimul and et al.}
}

@article{CHEN2025126215,
title = {Embedding dynamic graph attention mechanism into Clinical Knowledge Graph for enhanced diagnostic accuracy},
journal = {Expert Systems with Applications},
volume = {267},
pages = {126215},
year = {2025},
issn = {0957-4174},
doi = {https://doi.org/10.1016/j.eswa.2024.126215},
url = {https://www.sciencedirect.com/science/article/pii/S0957417424030823},
author = {Deng Chen and Weiwei Zhang and Zuohua Ding}
}

@misc{zuo2025kg4diagnosishierarchicalmultiagentllm,
      title={KG4Diagnosis: A Hierarchical Multi-Agent LLM Framework with Knowledge Graph Enhancement for Medical Diagnosis}, 
      author={Kaiwen Zuo and Yirui Jiang and Fan Mo and Pietro Lio},
      year={2025},
      eprint={2412.16833},
      archivePrefix={arXiv},
      primaryClass={cs.AI},
      url={https://arxiv.org/abs/2412.16833}, 
}

@article{Shakil_2024,
   title={Abstractive text summarization: State of the art, challenges, and improvements},
   volume={603},
   ISSN={0925-2312},
   url={http://dx.doi.org/10.1016/j.neucom.2024.128255},
   DOI={10.1016/j.neucom.2024.128255},
   journal={Neurocomputing},
   publisher={Elsevier BV},
   author={Shakil, Hassan and Farooq, Ahmad and Kalita, Jugal},
   year={2024},
   month=oct, pages={128255} }

@article{VanVeen_2023,
   title={Clinical Text Summarization: Adapting Large Language Models Can Outperform Human Experts},
   journal={Res Sq},
   year={2023},
   month={Oct},
   doi={10.21203/rs.3.rs-3483777/v1},
   author={Van Veen, D. and Van Uden, C. and Blankemeier, L. and et al.},
   note={Preprint, rs.3.rs-3483777}
}

@article{Adams_2023,
   title={Leveraging GPT-4 for post hoc transformation of free-text radiology reports into structured reporting: A multilingual feasibility study},
   journal={Radiology},
   volume={307},
   number={4},
   pages={e230725},
   year={2023},
   doi={https://doi.org/10.1148/radiol.230725},
   author={Adams, L.C. and Truhn, D. and Busch, F. and Kader, A. and Niehues, S.M. and Makowski, M.R. and Bressem, K.K.}
}

@misc{lou2024poweroptimizationintegratedactive,
      title={Power Optimization for Integrated Active and Passive Sensing in DFRC Systems}, 
      author={Xingliang Lou and Wenchao Xia and Kai-Kit Wong and Haitao Zhao and Tony Q. S. Quek and Hongbo Zhu},
      year={2024},
      eprint={2402.11294},
      archivePrefix={arXiv},
      primaryClass={cs.IT},
      url={https://arxiv.org/abs/2402.11294}, 
}

@misc{nori2023capabilitiesgpt4medicalchallenge,
      title={Capabilities of GPT-4 on Medical Challenge Problems}, 
      author={Harsha Nori and Nicholas King and Scott Mayer McKinney and Dean Carignan and Eric Horvitz},
      year={2023},
      eprint={2303.13375},
      archivePrefix={arXiv},
      primaryClass={cs.CL},
      url={https://arxiv.org/abs/2303.13375}, 
}

@article{Singhal_2023,
   title={Large language models encode clinical knowledge},
   journal={Nature},
   volume={620},
   pages={172--180},
   year={2023},
   doi={https://doi.org/10.1038/s41586-023-06291-2},
   author={Singhal, K. and Azizi, S. and Tu, T. and et al.}
}

@misc{liu2023visualinstructiontuning,
      title={Visual Instruction Tuning}, 
      author={Haotian Liu and Chunyuan Li and Qingyang Wu and Yong Jae Lee},
      year={2023},
      eprint={2304.08485},
      archivePrefix={arXiv},
      primaryClass={cs.CV},
      url={https://arxiv.org/abs/2304.08485}, 
}

@misc{wang2024yolov9learningwantlearn,
      title={YOLOv9: Learning What You Want to Learn Using Programmable Gradient Information}, 
      author={Chien-Yao Wang and I-Hau Yeh and Hong-Yuan Mark Liao},
      year={2024},
      eprint={2402.13616},
      archivePrefix={arXiv},
      primaryClass={cs.CV},
      url={https://arxiv.org/abs/2402.13616}, 
}

@misc{guo2024llavaultralargechineselanguage,
      title={LLaVA-Ultra: Large Chinese Language and Vision Assistant for Ultrasound}, 
      author={Xuechen Guo and Wenhao Chai and Shi-Yan Li and Gaoang Wang},
      year={2024},
      eprint={2410.15074},
      archivePrefix={arXiv},
      primaryClass={cs.CV},
      url={https://arxiv.org/abs/2410.15074}, 
}

@article{Li_2025,
   title={Intelligent detection and grading diagnosis of fresh rib fractures based on deep learning},
   journal={BMC Med Imaging},
   volume={25},
   number={1},
   pages={98},
   year={2025},
   month={Mar},
   doi={10.1186/s12880-025-01641-0},
   author={Li, T. and Liao, M. and Fu, Y. and et al.}
}

@inproceedings{Redmon_2016,
   title={You Only Look Once: Unified, Real-Time Object Detection},
   booktitle={Proceedings of the IEEE Conference on Computer Vision and Pattern Recognition (CVPR)},
   pages={779--788},
   year={2016},
   month={Jun},
   author={Redmon, J. and Divvala, S. and Girshick, R. and Farhadi, A.},
   address={Las Vegas, NV, USA}
}

@article{LeCun_1998,
   title={Gradient-Based Learning Applied to Document Recognition},
   journal={Proceedings of the IEEE},
   volume={86},
   number={11},
   pages={2278--2324},
   year={1998},
   month={Nov},
   author={LeCun, Y. and Bottou, L. and Bengio, Y. and Haffner, P.}
}

@article{Sun_2025,
   title={AI-assisted radiologists vs. standard double reading for rib fracture detection on CT images: A real-world clinical study},
   url={https://doi.org/10.1371/journal.pone.0316732},
   DOI={10.1371/journal.pone.0316732},
   journal={PLoS ONE},
   volume={20},
   number={1},
   pages={e0316732},
   year={2025},
   author={Sun, L. and Fan, Y. and Shi, S. and Sun, M. and Ma, Y. and Zhang, K. and et al.}
}

@article{FLORY2024152,
title = {Artificial Intelligence in Radiology: Opportunities and Challenges},
journal = {Seminars in Ultrasound, CT and MRI},
volume = {45},
number = {2},
pages = {152-160},
year = {2024},
note = {Perspectives in Radiology Practice: Challenges and Opportunities},
issn = {0887-2171},
doi = {https://doi.org/10.1053/j.sult.2024.02.004},
url = {https://www.sciencedirect.com/science/article/pii/S0887217124000052},
author = {Marta N. Flory and Sandy Napel and Emily B. Tsai}
}

@article{LIU2022283,
title = {Diagnostic value and limitations of CT in detecting rib fractures and analysis of missed rib fractures: a study based on early CT and follow-up CT as the reference standard},
journal = {Clinical Radiology},
volume = {77},
number = {4},
pages = {283-290},
year = {2022},
issn = {0009-9260},
doi = {https://doi.org/10.1016/j.crad.2022.01.035},
url = {https://www.sciencedirect.com/science/article/pii/S0009926022000435},
author = {C. Liu and Z. Chen and J. Xu and G. Wu},
}

@article{Chapman_2016,
   title={Clinical Utility of Chest Computed Tomography in Patients with Rib Fractures},
   url={http://dx.doi.org/10.5812/atr.37070},
   DOI={10.5812/atr.37070},
   journal={Arch Trauma Res},
   volume={5},
   number={4},
   pages={e37070},
   year={2016},
   author={Chapman, B.C. and Overbey, D.M. and Tesfalidet, F.}
}

@misc{openai2024gpt4,
      title={GPT-4 Technical Report}, 
      author = {{OpenAI Research Team}},
      year={2024},
      eprint={2303.08774},
      archivePrefix={arXiv},
      primaryClass={cs.CL}
}
\end{document}